\documentclass[10pt]{article}  
\usepackage{graphicx}
\usepackage{amssymb}
\input epsf
\parskip=\medskipamount
\def\k{\kappa}  
   
\def\om{\omega}   
\def\IB{\relax{\rm l\kern-.18 em B}}
\def\IC{\relax{\rm l\kern-.50 em C}}
\def\IE{\relax{\rm l\kern-.12 em E}}
\def\IK{\relax{\rm l\kern-.18 em K}}
\def\IL{\relax{\rm I\kern-.18 em L}}
\def\IN{\relax{\rm I\kern-.18 em N}}
\def\IR{\relax{\rm I\kern-.18 em R}}
\def\frac#1#2{{#1\over #2}}

\def\ptos{\leaders\hbox to 2mm{\hfil{.}\hfil}\hfill}
\def\\{\hfill\break}

\def\<#1>{\langle#1\rangle}

\def\Cos{\mathop{\rm C}\nolimits}    % funcion Coseno
\def\Sin{\mathop{\rm S}\nolimits}    % funcion Seno
\def\Tan{\mathop{\rm T}\nolimits}    % funcion Tangente
\def\k{\kappa}                       % kappa
\font\tenfrak=eufm10  \font\sevenfrak=eufm7  \font\fivefrak=eufm5
\newfam\frakfam
\textfont\frakfam=\tenfrak\scriptfont\frakfam=\sevenfrak
\scriptscriptfont\frakfam=\fivefrak

\font\tengoth=eufm10 scaled\magstep1 \font\sevengoth=eufm7
\font\fivegoth=eufm5
\newfam\gothfam
\textfont\gothfam=\tengoth\scriptfont\gothfam=\sevengoth
   \scriptscriptfont\gothfam=\fivegoth
\newtheorem{proposicion}{Proposition}

\begin{document}

\title{The Tremblay-Turbiner-Winternitz  system on spherical and hyperbolic spaces : Superintegrability, curvature-dependent formalism and complex factorization}

\author{ Manuel F. Ra\~nada  \\ [3pt]
{\sl Dep. de F\'{\i}sica Te\'orica and IUMA } \\
  {\sl Universidad de Zaragoza, 50009 Zaragoza, Spain}     }
%%  \date{ Mon, 17 Mar 2014}
%%  \date{}
%-------------------------------------------------------------------
%%  Version vRf  
%%  Mon, 17 Mar 2014
%-------------------------------------------------------------------
\maketitle

\begin{abstract}
The higher-order superintegrability of the Tremblay-Turbiner-Winternitz system (related to the harmonic oscillator) is studied on the two-dimensional spherical and hiperbolic spaces, $S_\k^2$ ($\k>0$),   and   $H_{\k}^2$ ($\k<0$). The curvature $\kappa$ is considered as a parameter and all the results are formulated in explicit dependence of $\kappa$. 
The idea is that the additional constant of motion can be factorized as the product of  powers  of two particular rather simple complex functions (here denoted by $M_r$ and $N_\phi$).  This technique leads to a proof of the superintegrability of the 
Tremblay-Turbiner-Winternitz system on  $S_\k^2$ ($\k>0$) and $H_{\k}^2$ ($\k<0$),   and to the explicit expression of the  constants of motion.
\end{abstract}

\begin{quote}
%----------------
 Integrability on spaces of constant curvature. Superintegrability. Higher-order  constants of motion. Complex factorization. 

{\sl Running title:}{\enskip}
The TTW system on spaces of constant curvature.

%----------------
%%  37J35   Completely integrable systems, topological structure
%%  70H06  Completely integrable systems and methods of integration

%----------------
%%  02.30.Ik   Integrable systems
%%  05.45.-a   Nonlinear dynamics and chaos
%%  45.20.Jj   Lagrangian and Hamiltonian mechanics
\end{quote}

\vfill
%----------------
\footnoterule{\small
\begin{quote}
  {\tt E-mail: {mfran@unizar.es}  }
\end{quote}
}

%----------------
\newpage
%----------------

%-----------------------------------------------------
%%  Section 1
\section{Introduction}

It is well known that systems that admit  Hamilton-Jacobi (or Schr\"odinger in the quantum case) separability in more than one coordinate system are superintegrable with  quadratic in the momenta constants of motion (in some particular cases the constant is determined by an exact Noether symmetry and then it is linear). For example, the following potential, known as the Smorodinsky-Winternitz (SW) potential \cite{FrMS65}--\cite{GroPoSi95}, and representing a two dimensional isotonic oscillator  \cite{WeiJo79}-\cite{Zhu87}, 
\begin{equation}
   V_{sw}  =  \frac{1}{2}\,\om_0^2(x^2 + y^2)  + \frac{k_2}{x^2}  + \frac{k_3}{y^2} \,,
\end{equation}
is separable in Cartesian and polar coordinates and it is, therefore,  superintegrable with three quadratic constants of motion 
(see \cite{MiPWJPa13} for a recent review on superintegrability). 

The potential  $V_{sw}$ admits two generalizations. The first one
\begin{equation}
  V(n_x,n_y)  =  \frac{1}{2}\,{\om_0}^2(n_x^2 x^2 + n_y^2 y^2)
    + \frac{k_1}{2x^2}  + \frac{k_2}{2y^2}  \,,  \label{Vn1n2}
\end{equation}
that preserves the separability in Cartesian coordinates,  is also superintegrable \cite{EvVe08}--\cite{RaRoS10} but with a polynomial of higher order than 2 as a third integral of motion. 
The second generalization of $V_{sw}$, that takes the form 
\begin{equation}
  V_{ttw}(r,\phi)  =  \frac{1}{2}\,{\om_0}^2 r^2 +  \frac{1}{2\,r^2}\,\Bigl(   \frac{\alpha}{\cos^2(m\phi)}  +  \frac{\beta}{\sin^2(m\phi)} \Bigr) \,,   \label{VttwE2}
\end{equation}
was firstly  studied by Tremblay, Turbiner, and Winternitz \cite{TTW09}--\cite{TTW10}, and then by other authors \cite{Qu10a}--\cite{CeKuNedO13}. 
When $m=1$ it reduces to $V_{sw}$, but in the general $m\ne 1$ case 
($m$ must be an integer or rational number)
it  is only separable in polar coordinates; therefore, the third integral is not quadratic in the momenta but a polynomial  of higher order than two (the degree of the polynomial depends of the value of $m$). 

The idea that the harmonic oscillator (and also the Kepler problem) can be correctly defined on spaces of constant curvature appears in a  book of Riemannian geometry of 1905 by  Liebmann \cite{Liebmann}; but it was Higgs \cite{Hi79} who studied this system with detail  (the study of Higgs was limited to a spherical geometry but his approach can be extended, introducing the appropriate changes,  to the hyperbolic space). The  TTW system is directly related to the harmonic oscillator; so it seems natural to also study the TTW system on constant curvature spaces. Actually, this question has been recently considered in \cite{ChDgR11} (TTW system but without the harmonic potential part) and in \cite{GonKas13ArXiv} (action-angle variables and perturbation theory).

The aim of this paper is to study the TTW system on the spaces of constant curvature $S_\k^2$  ($\k>0$) and $H_\k^2$  ($\k<0$), and to prove the superintegrability for all the values of $\kappa$. Two important points are: 

%----------------
\begin{itemize}

\item[ (a)]  All the mathematical expressions will depend of the curvature $\k$ as a parameter, in such a way that considering values $\k>0$, $\k=0$, or $\k<0$, we will obtain the corresponding property particularized for the system on the sphere $S_\k^2$, on the Euclidean space $\IE^2$,  or on the hyperbolic space $H_{\k}^2$, respectively. 
This curvature-dependent formalism was already used in \cite{RaSa02b}-\cite {RaSa03} (and in \cite{CRS11}-\cite {CRS12} for the quantum oscillator); other papers making use of this $\k$-dependent formalism are \cite{DoZi91}-\cite{HeBa06Sigma}. 

\item[ (b)] It is well known that the two dimensional harmonic oscillator with rational quotient of frequencies admits an third integral. 
The important point is  that this additional integral can be obtained as the product of two simple complex functions \cite{JauHillPr40} (see also \cite{RaRoS10}). 
The superintegrability of the standard Euclidean TTW system was proved in \cite{Ra12b} by using this technique. Now, in this paper, we present a generalization of this method to the $\kappa\ne 0$ case. 
\end{itemize}

 The paper  is organized as follows. In section 2 we first introduce the $\k$-dependent formalism and then we study the superintegrability of the he harmonic oscillator and the S-W potential on spaces of constant curvature. 
In section 3 we prove the superintegrabilty of the Tremblay-Turbiner-Winternitz  system on spherical and hyperbolic spaces. 
Finally in section 4 we make some comments and we present some  open questions.

%-----------------------------------------------------
%%  Section 2
\section{The harmonic oscillator on spaces of constant curvature }

%-----------------------------------------------------
%%   Seccion 2.1
\subsection{$\k$-dependent formalism  }

On a two--dimensional Riemannian space $(M,g)$ (not neccesarily
of constant curvature) there are two distinguished types of 
coordinate systems, ``geodesic parallel" and  ``geo\-desic polar"
coordinates, that reduce to the familiar Cartesian $(x,y)$ and polar
coordinates $(r,\phi)$ on the Euclidean plane \cite{Kl78}. 
Here we only consider  the geodesic polar coordinates that are 
based on a point $O$ and a oriented geodesic $l_0$ through $O$.
For any point $P$ in some suitable neighborhood a point $O$ 
(that represents  the origin)  there is a unique geodesic $l$ joining 
$O$ and $P$. The geodesic polar coordinates $(r,\phi)$ of $P$ are 
the  distance $r$ between $O$ and $P$ measured along $l$, 
and the angle $\phi$ between $l$ and the positive ray $l_0$  measured at $O$. 
These coordinates are are singular at $O$ and $\phi$ is discontinuous on the positive ray of $l_0$.

In what follows we will make use of the following $\kappa$-dependent trigonometric-hyperbolic functions
%----------------
\begin{equation}
  \Cos_{\k}(x) = \cases{
  \cos{\sqrt{\k}\,x}       &if $\k>0$, \cr
  {\quad}  1               &if $\k=0$, \cr
  \cosh\!{\sqrt{-\k}\,x}   &if $\k<0$, \cr}{\qquad}
%----------------
  \Sin_{\k}(x) = \cases{
  \frac{1}{\sqrt{\k}} \sin{\sqrt{\k}\,x}     &if $\k>0$, \cr
  {\quad}   x                                &if $\k=0$, \cr
  \frac{1}{\sqrt{-\k}}\sinh\!{\sqrt{-\k}\,x} &if $\k<0$, \label{SkCk}\cr}
\end{equation}
%----------------
and $\Tan_\k(x) = \Sin_{\k}(x)/\Cos_{\k}(x)$ \cite{RaSa02b}-\cite{HeBa06Sigma}. 
Then the following $\k$-dependent expression
\begin{equation}
 ds_\k^2 = d r^2 + \Sin_\k^2(r)\,d{\phi}^2 \,,
\end{equation}
represents the expression, in geodesic polar coordinates $(r,\phi)$, of the
differential line element  on the spaces $(S_{\k}^2,\IE^2,H_{\k}^2)$
with constant curvature $\k$.
This metric reduces to
$$
 ds_1^2 =    d r^2 + (\sin^2 r)\,d{\phi}^2 \,,{\quad}
 ds_0^2 =    d r^2 + r^2\,d{\phi}^2 \,,{\quad}
 ds_{-1}^2 = d r^2 + (\sinh^2 r)\,d{\phi}^2\,,
$$
in the three particular cases of the unit sphere $\k=1$, Euclidean plane $\k=0$,
and `unit' Lobachewski plane $\k=-1$.

A general standard Lagrangian ($\k$-dependent kinetic term minus a
potential) has the following form
$$
 L(r, \phi, v_r, v_\phi; \k) =  \frac{1}{2}\,\Bigl(\,v_r^2 + \Sin_\k^2(r)
v_{\phi}^2\,\Bigr) -  U(r,\phi;\k)  \,,
$$
in such a way that for $\k=0$ we recover the  expression of a standard Lagrangian in the Euclidean space.
The two linear momenta, reducing to $p_x$ and $p_y$, in the Euclidean case, are given by 
\begin{eqnarray*}
 P_1(\k)  &=&  (\cos{\phi})\,v_r - (\Cos_\k(r) \Sin_\k(r)\sin{\phi})\,v_{\phi}  \cr
 P_2(\k)  &=& (\sin{\phi})\,v_r + (\Cos_\k(r) \Sin_\k(r)\cos{\phi})\,v_{\phi}
\end{eqnarray*}
and the $\k$-dependent expression for the angular momentum is
$$
 J(\k)  = \Sin_{\k}^2(r)\,v_{\phi} \,.
$$

%-----------------------------------------------------
%%   Seccion 2.2
\subsection{The harmonic oscillator on spaces of constant curvature } 

 The following (spherical, Euclidean, hyperbolic) Lagrangian with curvature $\k$,
\begin{equation}
 L(\k) = \frac{1}{2}\,\Bigl(\,v_r^2 + \Sin_\k^2(r) v_{\phi}^2\,\Bigr)
        -  U(r;\k) \,,{\quad}
 U(r;\k) = \frac{1}{2}\,\om_0^2\,\Tan_\k^2(r) \,,
\end{equation}
represents the $\k$-dependent version of the harmonic oscillator
\cite{RaSa02b,RaSa03}; the potential $U(r;\k)$ reduces to
$$
 U_1 =  \frac{1}{2}\,\om_0^2\,\tan^2\!r  \,,{\quad}
 U_0 = V = \frac{1}{2}\,\om_0^2\,r^2  \,,{\quad}
 U_{-1} = \frac{1}{2}\,\om_0^2\,\tanh^2\!r \,,
$$
in the three particular cases of the unit sphere ($\k=1$), Euclidean plane ($\k=0$), 
and `unit' Lobachewski plane ($\k=-1$); the Euclidean function $V(r)$ appears in this formalism as making separation between two different behaviours (see Figure 1); 
of course, the domain of $r$ depends of the value of $\k$; we have $r\in[0,\infty)$ for $\k\le 0$ and $r\in[0,\pi/2\sqrt{\k}]$ for $\k>0$. 
It is known \cite{RaSa02b}-\cite{RaSa03} that this system is superintegrable for all the values of the curvature $\k$ since that, in addition to the angular momentum
$J(\k)$, it is endowed with the following two quadratic constants of the motion
%----------------
\begin{eqnarray*}
 I_1(\k) &=&  P_1^2(\k) +  {\om_0^2}\,\bigl(\Tan_\k(r) \cos{\phi}\bigr)^2 \,,\cr
 I_2(\k) &=&  P_2^2(\k) +  {\om_0^2}\,\bigl(\Tan_\k(r) \sin{\phi}\bigr)^2 \,,  
\end{eqnarray*}
%----------------
in such a way hat the energy can be written as follows
$$
 E(\k) = \frac{1}{2}\,\Bigl(\,  I_1(\k) + I_2(\k) + \k J^2(\k) \,\Bigr)  \,.
$$
An additional interesting property is the existence of the
following fourth integral of motion
$$
 I_4(\k) = P_1(\k)P_2(\k) + {\om_0^2}\,\bigl( \Tan_\k^2(r) \cos{\phi}\sin{\phi} \bigr)  \,.
$$
The reason is that, although it is not functionally independent
since it satisfies the following relation
$$
 I_4^2(\k) = I_1(\k) I_2(\k) - \om_0^2 J^2(\k)  \,,
$$
the set of the three $\k$-dependent functions $\bigl\{
I_1(\k),I_2(\k),I_4(\k) \bigr\}$ can be considered as the three
components of the $\k$-dependent version of the Fradkin tensor \cite{Frad65}.

%-----------------------------------------------------
%%   Seccion 2.3
\subsection{The S-W potential on spaces of constant curvature }

 The following (spherical, Euclidean, hyperbolic)  $\k$-dependent potential 
\begin{equation}
 U(r,\phi;\k) =  \frac{1}{2}\,\om_0^2\,\Tan_\k^2(r) 
 +   \frac{k_2 }{(\Sin_\k(r)\cos{\phi})^2}  +\frac{k_3 }{(\Sin_\k(r)\sin{\phi})^2}  \,,
\end{equation}
that is well defined for all the values of $\k$, represents the spherical ($k>0$) and hyperbolic ($\k<0$) version of the Euclidean potential $V_{sw}$ ($\k=0$);  it reduces to
\begin{eqnarray}
 U_1  &=&  \frac{1}{2}\,\om_0^2\,\tan^2\!r + 
 \frac{1}{\sin^2 r} \Bigl(  \frac{k_2}{\cos^2{\phi}}  + \frac{k_3}{\sin^2{\phi}} \Bigr)\,,\cr
 U_{-1} &=&  \frac{1}{2}\,\om_0^2\,\tanh^2\!r + 
 \frac{1}{\sinh^2 r} \Bigl(  \frac{k_2}{\cos^2{\phi}}  + \frac{k_3}{\sin^2{\phi}} \Bigr)\,,
{\nonumber}
\end{eqnarray}
in the particular cases of the unit sphere ($\k=1$) and `unit' Lobachewski plane ($\k=-1$). It is endowed with the following three quadratic constants of the motion
%----------------
\begin{eqnarray*}
  I_1(\k) &=& P_1^2(\k)
           + {\om_0^2}\,\bigl(\Tan_\k(r)\,\cos{\phi}\bigr)^2 
           + \frac{2\,k_2}{\bigl(\Tan_\k(r)\,\cos{\phi}\bigr)^2}  \,,\cr 
  I_2(\k) &=& P_2^2(\k)
           + {\om_0^2}\,\bigl(\Tan_\k(r)\,\sin{\phi}\bigr)^2  
           + \frac{2\,k_3}{\bigl(\Tan_\k(r)\,\sin{\phi}\bigr)^2}  \,,\cr 
  I_3(\k) &=& J^2(\k) + \frac{2\,k_2}{\cos^2{\phi}}  
           + \frac{2\,k_3}{\sin^2{\phi}} \,.   
\end{eqnarray*}
%--------------- 
and, therefore, it is a superintegrable system for all the values of $\k$.

%-----------------------------------------------------
%%  Section 3  
\section{The TTW system on spaces of constant curvature }

In the following, we will make use of the Hamiltonian formalism; therefore, the time 
derivative $d/dt$ of a function means the Poisson bracket of the function with the Hamiltonian.

We have seen, in the previous section 2, that in both the harmonic oscillator and the S-W potential the curvature $\kappa$ modify many things but preserve the fundamental property of superintegrability. Now in this section we will prove that this is also true for the TTW system

It is well known that  if $F(\phi)$ are arbitrary function then the following Hamiltonian 
(harmonic oscillator plus an angular deformation introduced by $F$) 
\begin{equation}
 H =   \frac{1}{2}\,\bigl(p_r^2 + \frac{p_\phi^2}{r^2}\bigr) + 
 \frac{1}{2}\,{\om_0}^2 r^2 + \frac{1}{2}\,\frac{F(\phi)}{r^2} \,.  \label{H(rfi}
\end{equation}
is separable in in polar coordinates and  it is therefore endowed with the following  two constants of the motion 
\begin{eqnarray*}  
 J_1 &=& p_r^2 + \frac{p_\phi^2}{r^2} + {\om_0}^2 r^2 +\frac{F(\phi)}{r^2}  \cr 
 J_2 &=& p_\phi^2 + F(\phi)
\end{eqnarray*}

The following propsition states this property for spherical ($\k>0$) and hyperbolic ($\k<0$) spaces. 

%----------------
%%  (Proposicion 1)
\begin{proposicion}  \label{prop1}
The Hamiltonian
\begin{equation}
   H(\k)  =  \frac{1}{2}\,\Bigl(\, p_r^2 +  \frac{p_{\phi}^2}{\Sin_\k^2(r)}\,\Bigr) +  \frac{1}{2}\,\om_0^2\,\Tan_\k^2(r) + \frac{1}{2}\,\frac{F(\phi)}{(\Sin_\k(r))^2}  \label{Hk}
\end{equation}
is separable in geodesic polar coordinates $(r,\phi)$ and  it is endowed with the following  two constants of the motion 
\begin{eqnarray*}  
 J_1 &=& p_r^2 + \frac{p_{\phi}^2}{\Sin_\k^2(r)} +  \om_0^2\,\Tan_\k^2(r) + \frac{F(\phi)}{(\Sin_\k(r))^2}  \cr 
 J_2 &=& p_\phi^2 + F(\phi)
\end{eqnarray*}
 This property is true for all the values of the curvature $\k$. 
\end{proposicion}

As we comment in the introduction, the TTW system is separable in the Euclidean plane in polar coordinates. Now we see that it admits a generalization to the spaces $S_\k^2$  ($\k>0$) and $H_\k^2$  ($\k<0$) that appears as a particular case of the Hamiltonian (\ref{Hk}); therefore, it is also separable (and therefore integrable) in  spherical and hyperbolic spaces. 

The following proposition proves the superintegrability of the TTW system on spaces of constant curvature and presents a method for obtaining the explicit expression of the third integral of motion. 
%----------------
%%  (Proposicion 2)
\begin{proposicion}  \label{prop2}
Consider the nonlinear harmonic oscillator-related potential
\begin{equation}
  U_{m}(r,\phi)  =   \frac{1}{2}\,\om_0^2\,\Tan_\k^2(r) + \frac{1}{2}\,\frac{F_m(\phi)}{(\Sin_\k(r))^2}   \,,{\quad}
  F_m(\phi) = \frac{k_a} {\sin^2(m\phi)} +  k_b\,\Bigl(\frac {\cos(m\phi)} {\sin^2(m\phi)}\Bigr) \,,  
   \label{UkFm}
\end{equation}
where $k_a$ and $k_b$ are arbitrary constants.
Let $J_1$ and $J_2$ the two quadratic constants of motion associated to the Liouville integrability 
\begin{eqnarray*}  
 J_1 &=& p_r^2 + \frac{p_{\phi}^2}{\Sin_\k^2(r)} +  \om_0^2\,\Tan_\k^2(r) + \frac{F_m}{\Sin_\k^2(r)}  \cr 
 J_2 &=& p_\phi^2 + F_m
\end{eqnarray*}
and let $M_r$ and $N_\phi$ be the  complex functions   $M_r = M_{r1} + i\,  M_{r2}$ and  $N_{\phi} = N_{\phi 1} +  i\,   N_{\phi 2}$ with real and imaginary parts, $M_{r a}$ and $N_{\phi a}$, $a=1,2$, be defined as 
$$
 M_{r1} =  \frac{2}{\Tan_\k(r)}\,p_r\,\sqrt{J_2} \,,{\qquad}
 M_{r2} =   p_r^2 + \om_0^2 \,\Tan_\k^2(r) - \frac{ J_2}{\Tan_\k^2(r)}
 = J_1 - \frac{1 + \Cos_\k^2(r)}{\Sin_\k^2(r)} \, J_2  \,,
$$
$$
 N_{\phi 1} =  \frac{k_b}{2} +  J_2\,\cos(m\phi)  \,,{\qquad}
 N_{\phi 2} =   \sqrt{J_2} \,p_\phi\,\sin(m\phi) \,.
$$ 
Then, the complex function $K_m$ defined as
$$
  K_m = M_r^{m} (N_\phi^{*})^{2}
$$
is a (complex) constant of the motion.  
\end{proposicion}
{\it Proof:} 
First, let us comment that the functions $M_{r 1}$ and $M_{r 2}$ are $\k$-dependent but they satisfy the appropriate Euclidean limit \cite{Ra12b}
$$
  \lim_{\k\to 0} M_{r1}  =  \frac{2}{r}\,p_r\,\sqrt{J_2} \,,{\qquad}
  \lim_{\k\to 0} M_{r2}  =  p_r^2 + \om_0^2 r^2 - \frac{ J_2}{r^2} 
  = J_1 - \frac{2}{r^2}J_2  \,.
$$
The expresions of the functions $N_{\phi 1}$ and $N_{\phi 2}$ are the same as in the Euclidean plane.

The time-derivative (Poisson bracket with $H(\kappa)$) of the function $M_{r1}$ is proportional to $M_{r2}$ and the time-derivative of the $M_{r2}$ is proportional to $M_{r1}$ but with the opposite sign 
$$
 \frac{d}{d t}\,M_{r1}=  -\,2\,{\lambda_\k}\,M_{r2} \,,{\quad}
 \frac{d}{d t}\,M_{r2} =  2 {\lambda_\k}\,M_{r1}  \,,
$$
and this property is also true for the angular functions 
$$
 \frac{d}{d t}\,N_{\phi 1} =  -\,m\,{\lambda_\k}\,N_{\phi 2} \,,{\quad}
 \frac{d}{d t}\,N_{\phi 2} =  m \,{\lambda_\k}\,N_{\phi 1}  \,,{\quad} 
$$
where the common factor ${\lambda_\k}$ takes the value 
$$
 {\lambda_\k} = \frac{1}{\Sin_\k^2(r)}\,\sqrt{J_2}  \,,{\quad}
  {\lambda_0} = \frac{1}{r^2}\,\sqrt{J_2}   \,. 
$$
Therefore, the time-evolution of the complex functions $M_r$ and $N_\phi$ is given by
$$
 \frac{d}{d t}\,M_r  =  i\,   2 {\lambda_\k}\,M_r  \,,{\quad}
 \frac{d}{d t}\,N_\phi  =  i\,   m\,{\lambda_\k}\,N_\phi   \,,{\quad}
$$
Thus we have
%----------------
\begin{eqnarray*}
  \frac{d}{dt}\,K_m &=&  \frac{d}{dt}\,\Bigl( M_r^{m} (N_\phi^{*})^{2}\Bigr) = M_r^{(m-1)}N_\phi^{*}\,\Bigl( \, m\,\dot{M_r}\,N_\phi^{*}   +  2 M_r\,\dot{N_\phi}^{*} \,\Bigr)   \cr
  &=& M_r^{(m-1)}N_\phi^{*}\,\Bigl( \, m\, i\,  2 {\lambda_\k}\,M_r\,N_\phi^{*}   +  2 M_r\,(- i\,  m\,{\lambda_\k}\,N_\phi^{*}) \,\Bigr) =  0  \,.
\end{eqnarray*}

Finally, let us comment that the moduli of these two complex functions (that are constant of the motion of fourth order in the momenta)  are given by
\begin{eqnarray*}
\mid M_r \mid^2  &=& 4(H^2 - \om_0^2 J_2) + \k (\k J_2 - 4 H) J_2 \cr
\mid N_\phi \mid^2 &=&  J_2^2 - k_a J_2 + \frac{k_b^2}{4}
\end{eqnarray*}
\hfill${\square}$

Summarizing: the TTW is super-integrable for any value of the curvature (positive, zero or negative) and the additional constant of motion $K_m$ can be obtained by 
complex factorization. Since the function $K_m$ is complex it can be written as $K_m=J_3+  i\,  J_4$ with $J_3$ and $J_4$ real constants of the motion, that is, $dJ_3/dt=0$, $dJ_4/dt=0$. One of them, for example $J_3$, can be chosen as the third fundamental integral of the motion. 

 The function $F_m(\phi)$ in (\ref{UkFm}) can be considered, at first sight, as not the same as the angular function in the original TTW  potential (\ref{VttwE2}). Nevertheless it can be proved the following trigonometric equality 
 $$
  \frac{2(\alpha+\beta)} {\sin^2(2m\phi)} +  2(\beta - \alpha)\,\Bigl(\frac {\cos(2 m\phi)} {\sin^2(2m\phi)}\Bigr) =  \frac{\alpha} {\cos^2(m\phi)} +  \frac {\beta} {\sin^2(m\phi)}  \,.   
$$
Thus, the above proposition \ref{prop2} is also true for the potential $U_{m}$ rewritten with the angular function as in (\ref{VttwE2}).
More specifically, let us now consider the following (spherical, Euclidean, hyperbolic) potential  
\begin{equation}
  U_m'(r,\phi)  =   \frac{1}{2}\,\om_0^2\,\Tan_\k^2(r) + \frac{1}{2}\,\frac{G_m(\phi)}{\Sin_\k^2(r)}     \,,{\quad}
  G_m(\phi) = \frac{\alpha} {\cos^2(m\phi)} +  \frac {\beta} {\sin^2(m\phi)} \,.     
    \label{UkGm}
\end{equation}
and let us denot denote by $J_1'$ and $J_2'$ the two constants of motion $J_1$ and $J_2$ but now rewritten as functions of $G_m$ 
\begin{eqnarray*}  
 J_1' &=& p_r^2 + \frac{p_{\phi}^2}{\Sin_\k^2(r)} +  \om_0^2\,\Tan_\k^2(r) + \frac{G_m}{\Sin_\k^2(r)}   \cr 
 J_2' &=& p_\phi^2 + G_m
\end{eqnarray*}
Then if we also write with primes the new functions $M_r$ and $N_\phi$ 
$$
  M_{r1}' =  \frac{2}{\Tan_\k(r)}\,p_r\,\sqrt{J_2'} \,,{\qquad}
  M_{r2}' =   p_r^2 + \om_0^2 \,\Tan_\k^2(r) - \frac{J_2'}{\Tan_\k^2(r)} \,, 
$$
$$
 N_{\phi 1}' =  \beta-\alpha+  J_2'\,\cos(2 m\phi)  \,,{\qquad}
 N_{\phi 2}' =   \sqrt{J_2'} \,p_\phi\,\sin(2 m\phi) \,, 
$$ 
the complex constant of motion for the potential (\ref{UkGm}) is now given by 
$$
  K_m' = ({M_r'})^{2 m} ({N_\phi'}^{*})^{2}  \,. 
$$

%-----------------------------------------------------
%%  Section 4   
\section{Final comments  }

The following two points summarize the main results proved in this paper.
%----------------
\begin{itemize}
\item  The TTW system is not a specific characteristic of the Euclidean space but it is well defined in all the three spaces of constant curvature. 
Moreover, we have represented the TTW system by a unique Hamiltonian [with potential (\ref{UkFm})  or with potential (\ref{UkGm})] that is a smooth function of the curvature $\kappa$ and, in this way,  we can say that there are not three different TTW systems but only one that is defined, at the same time, in the three different manifolds.

\item The TTW system is superintegrable in the three spaces of constant curvature. The adidtional third integral of motion can be explicitly obtained as the product of  powers of two particular rather simple complex functions (here denoted by $M_r$ and $N_\phi$).  This factorization, that is valid for all the values of $\k$, generalizes the Euclidean property previously  proved in ref. \cite{Ra12b}. 
 
\end{itemize}

We conclude with the following two comments: First, it has been recently proved the superintegrabilty of another Euclidean system, known as the PW system, similar to the TTW but related with the Kepler problem \cite{PostWint10}-\cite{Ra13}. 
We think that the PW system can also be studied on spaces of constant curvature by making use of the curvature-dependent formalism.  
Second, the TTW system is also important at the quantum level. The properties of the functions $M_r$ and $N_\phi$ can probably be interesting  (changing functions for operators) for the study of the quantum Schr\"odinger equation by the method of factorization and ladder operators. 

%-----------------------------------------------------
\section*{Acknowledgments}

This work was supported by the research projects MTM--2012--33575 (MICINN, Madrid)  and DGA-E24/1 (DGA, Zaragoza).

%----------------
\begin{figure}\centerline{
%%  \epsfbox{QHOrphiFig1.eps}
\includegraphics{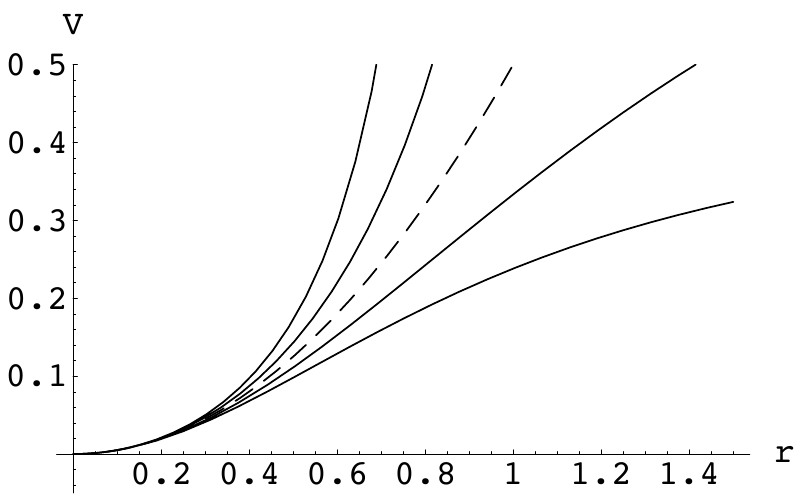}  }

\caption{Plot of the potential  $U(r,k)=(1/2)\,\om_0^2\,\Tan_\k^2(r)$, $\om_0=1$, as a function of $r$, 
for $\k<0$ (lower curves), $\k=0$ (dash line), and $\k>0$ (upper curves).}
\label{Fig1}
\end{figure}

%----------------
 \vfill\eject
%----------------

%----------------
\vfill\eject
%----------------

{\small
%-------------------
 }
%-------------------
%-------------------
\end{document}